\documentclass[final]{cvpr}

\usepackage{times}
\usepackage{epsfig}
\usepackage{graphicx}
\usepackage{amsmath}
\usepackage{amssymb}

\usepackage[breaklinks=true,colorlinks,bookmarks=false]{hyperref}
\begin{document}

%%%%%%%%% TITLE
\title{Interactions of Tris with rutile surfaces and consequences\\
for \textit{in vitro} bioactivity testing}

\author{\underline{Azade YazdanYar}\textsuperscript{1 %\href{ayazdan@icp.uni-stuttgart.de}{\textbardbl}
\thanks{Present address: Institute for Computational Physics, Universität Stuttgart, Allmandring 3, 70569 Stuttgart, Germany; ayazdan@icp.uni-stuttgart.de} }, 
Léa Buswell\textsuperscript{1}, 
Delphin Pantaloni\textsuperscript{1 \thanks{Present address: Université de Bretagne-Sud, IRDL, CNRS UMR 6027, BP 92116, 56321 Lorient Cedex, France}}, 
Ulrich Aschauer\textsuperscript{2}, 
Paul Bowen\vspace{4mm}\textsuperscript{1}\\

\small \textsuperscript{1}Department of Materials Science and Engineering, École Polytechnique Fédérale de Lausanne (EPFL), \\
\small Route Cantonale, Lausanne 1015, Switzerland\\
\small \textsuperscript{2}Department of Chemistry and Biochemistry, University of Bern, Freiestrasse 3, Bern 3012, Switzerland
}

\maketitle

%%%%%%%%% ABSTRACT
\begin{abstract}
Tris(hydroxymethyl)aminomethane (Tris) has been used as the buffer in bioactivity testing for over two decades and has become a standard choice for the scientific community. While it is believed to be non-interacting, the extent of its interactions with titanium oxide surfaces has not been systematically studied. Here, we use experimental (zeta potential measurements) and computational (molecular dynamics) approaches to evaluate the interaction of Tris with a rutile surface and how it affects the adsorption of other molecules relevant in biomedical \textit{in vitro} testing. We show that the interaction of Tris with the rutile surface is strong and significantly affects the interaction of other organic residues with the surface. These strong interactions are compounded by the Tris concentration in the \textit{in vitro} testing protocol which is much higher compared to other components. Our findings indicate that the kinetics observed in \textit{in vitro} tests will be strongly influenced by the presence of Tris as a buffering agent when compared to the natural CO\textsubscript{2} buffer in blood. These results reveal that considering the so-far neglected active role of Tris in \textit{in vitro} testing is critically needed and that \textit{in vitro} protocols using CO\textsubscript{2} partial pressure as the buffering agent should yield more reliable results.

% \noindent \textbf{Keywords}: Rutile; Tris(hydroxymethyl)aminomethane (Tris); Amino acid; \textit{In vitro} testing; Free energy calculations
\end{abstract}

%%%%%%%%% BODY TEXT
\section{Introduction}
\noindent In many bio-related applications, biomaterial surfaces and their properties are widely studied since they affect or determine the interaction of the surrounding environment with the biomaterial. A thorough understanding of the nature and mechanism of these interactions can help the scientific community to design and propose surfaces with enhanced and targeted bio-properties and performances.

Prior to commercial application, candidate biomaterials are tested using \textit{in vivo}, \textit{in vitro} and computational approaches to ensure their performance and reliability. Although the most accurate results are obtained by \textit{in vivo} methods, since testing conditions are the closest to the final working conditions, their complexity and need of expertise, rather high cost and sacrifice of living animals, often drive the researchers to prefer \textit{in vitro} and computational methods.

Titanium-based biomaterials, widely used as biomedical implants \cite{diebold_surface_2003, muller_small-diameter_2015,kokubo_novel_2003,liu_surface_2004,jager_significance_2007,qian_selective_2014}, are no exception. However, many \textit{in vivo} tests are still performed for bioactivity testing, the main reason being the unsatisfactory number of false positive and false negative results from \textit{in vitro} tests \cite{hulsart-billstrom_surprisingly_2016, zadpoor_relationship_2014}. While in the following we will discuss possible explanations for discrepancies between \textit{in vivo} and \textit{in vitro} results, our primary goal is to warn the scientific community of possible shortcomings with current \textit{in vitro} testing standards and to motivate the development of refined methodologies rather than relying on undesirable \textit{in vivo} testing. \textit{in vivo} testing should be performed only when necessary and after extensive and rigorous \textit{in vitro} testing due to ethical concerns and animal welfare regulations \cite{hubrecht_3rs_2019}.

Current \textit{in vitro} testing protocols for titanium-based biomaterials \cite{noauthor_iso_2007} use an ionic solution called Simulated Body Fluid (SBF), and considers the formation of hydroxyapatite, a calcium phosphate mineral, on the surface of the sample as a positive sign of bioactivity. Recent work \cite{baino_use_2020} reviews the effect of different parameters for \textit{in vitro} bioactivity testing using SBFs. Although the ionic composition of SBF is very close to human blood plasma \cite{kokubo_novel_2003}, in which \textit{in vivo} tests are performed, there are differences between the two that may cause false negative and false positive results. Bohner and Lemaitre \cite{bohner_can_2009} highlighted the limitations of using SBFs in current bioactivity testing protocols since i) contrary to blood serum, it does not contain any proteins, ii) there is no control on the carbonate content in SBF despite the fact that carbonates acts as the pH buffer in blood serum, and finally, iii) tris(hydroxymethyl)aminomethane (Tris), is used to buffer the pH at 7.4.

Zhao \etal. \cite{zhao_comparative_2017} addressed the significance of these differences in their more recent study where they compared a Tris buffered SBF and carbonate buffered SBF in absence and presence of bovine serum albumin (BSA). They observed that hydroxyapatite formation on the test surface was significantly inhibited at BSA concentrations much lower than that of proteins in blood plasma. The use of the carbonate buffer resulted in more reliable results than experiments using the Tris buffer \cite{zhao_comparative_2017}. This lead the authors to recommend the use of CO\textsubscript{2} instead of Tris, which is not a natural buffer and is not present in blood plasma.

Tris is proposed as the buffering agent in the ISO standard \cite{noauthor_iso_2007} and is assumed to be non-interacting. However, given its very high concentration (974 mM) in the ISO standard \cite{noauthor_iso_2007}, one may expect that it could have an influence if it exhibits any affinity for the biomaterial surface. Tris may also affect the interaction between blood plasma components and the surface of the test biomaterial and modify its bioactivity. Indeed, it was already shown that Tris strongly interacts with organic molecules such as peptide backbones, BSA \cite{taha_interactions_2010} and lysozyme \cite{quan_resurveying_2008, salis_not_2016, kang_visualization_2014}.

Rutile is present in heat-treated, as well as chemically-treated titanium surfaces used as medical implants \cite{liu_surface_2004, qian_selective_2014,kim_hyunmin_preparation_1996, kokubo_novel_2016, li_current_2011, roach_tuning_2016,carrado_nanoporous_2017} and is often used as a model surface for the interaction of different species found in SBFs and blood plasma \cite{zhao_comparative_2017,kokubo_novel_2016,yazdanyar_interaction_2018, zhao_rapid_2019,anselme_osteoblast_2000, tas_use_2014}. Tris was previously shown to interact with titanium dioxide surfaces via FTIR spectroscopy \cite{loreto_effect_2017}. In another study on the bioactivity evaluation of glass-ceramic scaffolds, the dissociation rate of the glass-ceramic scaffold was doubled in Tris-buffered solutions. It was also observed that presence or absence of Tris determines the crystalline phase of the apatite, that developed on the scaffold surface \cite{rohanova_tris_2011}.

The above findings, motivate the question of the extent of Tris interactions with the rutile surface, and its effect on the interaction of amino acids with the surface. Consequently, here we make a simple pragmatic investigation on the interaction of Tris with a model rutile surface using both experimental (zeta potential measurement) and computational (molecular dynamics) approaches. In particular, we study the effect of Tris on amino acid interactions with the rutile surface via zeta potential measurements to ascertain how competitive Tris may be with other molecules expected to adsorb on the rutile surface. Computational simulations complement our previous study on the interaction of a series of amino acids with a model rutile (110) surface \cite{yazdanyar_adsorption_2018} and support the experimental findings of a significant suppression of amino acids adsorption on rutile by Tris. Despite growing concerns regarding the reliability of Tris-buffered SBF solutions for \textit{in vitro} testing of biomaterials, and the effort of various groups to present refined protocols \cite{zhao_rapid_2019}, \textit{in vitro} bioactivity tests are still very commonly performed using Tris \cite{ballarre_versatile_2020,ferraris_antibacterial_2014,pereira_titanium_2014,el-wassefy_assessment_2014,zhang_preparation_2016, yao_effects_2019,todea_effect_2019,ren_morphologically_2018,liu_enhanced_2019,catauro_coatings_2016,ferraris_micro-_2015,dominguez-trujillo_bioactive_2018,caparros_bioactive_2016,yamaguchi_two--one_2017,kalaivani_effect_2014,durdu_bioactive_2016,gao_dopamine_2021,sankaralingam_preparation_2021,jariya_development_2021,stevanovic_assessing_2020,ahuja_bioactivity_2020,coelho_biomimetic_2020}. Our results highlight and quantify the so-far mostly overlooked role of Tris and should motivate the scientific community to establish \textit{in vitro} testing protocols that are free of Tris-induced artefacts.
%-------------------------------------------------------------------------
\section{Materials and Methods}
\subsection{Experimental methods}
\subsubsection{Rutile suspension and Tris solution}
Rutile powder with a purity of 99.99\% from Sigma-Aldrich was used. A surface area of 2.48 $\pm$ 0.02 m\textsuperscript{2}/g and mean diameter, D\textsubscript{v}[4,3] of 8.09 $\mu$m were determined using nitrogen adsorption (BET model, Gemini 2375, Micromeritics Instrument, Norcross, GA, USA) and laser diffraction (Malvern Mastersizer S, Malvern Instruments Ltd., Worcestershire, UK), respectively. The X-ray diffraction (XRD) pattern of the powder showed that rutile has negligible traces of anatase (Figure S1, Philips X Pert, Eindhoven, The Netherlands). No trace of organic contamination was found in the thermogravimetric analysis carried out in air (TGA, TGA/SDTA851e, Mettler Toledo, Columbus, OH, USA).

The TiO\textsubscript{2} suspension was prepared with 3 g of rutile powder and 157 g of either a solution of 10 mM NaCl with a purity of 99.5\%, or a Tris solution (10 mM NaCl and 50 mM Tris) in Ultrapure water. The rutile concentration of 1.875 wt\% was chosen according to the AcoustoSizer II (Colloidal dynamics, Florida, USA) sensitivity. Details of the suppliers and references for all the purchased chemicals can be found in Table S1.

The Tris solution was prepared at 25°C with a pH of 7.4, similar to the physiological pH. The suspension was dispersed using a magnetic agitator for 15 minutes, before zeta potential measurements using the AcoustoSizer II. These were performed at ambient temperature in the following manner for the different suspensions and components of interest.
%%%%%
\subsubsection{Acid-base titration of the rutile suspension}
The rutile suspension, dispersed in 10 mM NaCl, was titrated from pH 3 to 12. A pH of 3 was obtained using 1 M HCl and then 1 M NaOH was added automatically by the Acoustosizer II up to a pH of 12, to measure the evolution of the zeta potential as a function of the pH.
%%%%%
\subsubsection{Acid-base titration of the rutile suspension with Tris}
The acid-base titration was performed on rutile, dispersed in 10 mM NaCl and 50 mM Tris, at pH 7.4. First, 1 M HCl was added until a pH of 2 was obtained and then, 1 M NaOH was added automatically by the Acoustosizer II until a pH of 11 was obtained. 
%%%%%
\subsubsection{Tris titration of the rutile suspension}
To check for adsorption of Tris on the rutile surface, the rutile suspension in 10 mM NaCl was titrated automatically by the Acoustosizer II with the solution of 50 mM Tris at pH 7.4. This concentration is much lower than the Tris concentration in the \textit{in vitro} solution, which is 974 mM \cite{noauthor_iso_2007} but sufficient to buffer the pH of the suspension.
\subsubsection{Amino acids titration of the rutile suspension}
The amino acid solutions used to titrate the rutile suspension were prepared in ultrapure water with a concentration of 0.1 wt\%, which is lower than the concentration of plasma proteins (8 wt\%) \cite{marieb_human_2013}, in order to track the effect of very small amounts of amino acids on the zeta potential.

As each amino acid behaves differently, the pH was adjusted either with HCl or NaOH to obtain a pH of 8.5. HCl with a concentration of 5 mM was used for arginine, while NaOH with a concentration of 0.8, 1.3 and 9 mM was used for alanine, serine and aspartic acid, respectively. Tests were repeated several times. Details of the purchased amino acids can be found in Table S1. 

Titrations of the rutile suspension with and without Tris were performed to study the adsorption of two of the amino acids (Ala and Arg) on the rutile surface. 20 mL of 1 M NaOH was added to the suspensions without Tris to obtain a pH of 8.5. The pH was not adjusted for the experiments with Tris as it buffers the pH to 7.4.
%%%%%
\subsection{Computational methods}
\subsubsection{The rutile surface}
The simulation box was set up similar to our previous work \cite{yazdanyar_adsorption_2018}. The rutile (110) slab with a lateral dimension of 35.51×38.98 Å\textsuperscript{2}, and a thickness of 70 Å, was used. As rutile is hydroxylated in presence of water, the surface was first fully hydroxylated with 72 bridging and 72 terminal hydroxyl groups. A partial negative charge of -0.011 Cm\textsuperscript{-2} was then induced on the surface by deprotonating one bridging hydroxyl, hereafter referred to as the surface charge point, to mimic the negative charge of the rutile surface under physiological conditions. Details of different species of the rutile slab and their partial charges can be found in our previous work \cite{yazdanyar_adsorption_2018}.

A water layer with a thickness of 90 Å was added to the rutile slab in the z direction. Periodic boundary conditions were applied in all directions, with the periodic images of the simulation box in the z direction being separated from each other with a vacuum gap of 100 Å, in order to prevent interaction between replicas of the rutile slab in this direction.
%%%%%
\subsubsection{The Tris molecule}
Tris ((HOCH\textsubscript{2})\textsubscript{3}CNH\textsubscript{2}), has a $pK_a$ of 8.3 \cite{campbell_biochemistry_2014}. We are interested in the physiological pH of 7.4, at which the Tris molecule is in its protonated form of (HOCH\textsubscript{2})\textsubscript{3}CNH$_3^+$. The coordinates of the protonated molecule were taken from Ligandbook \cite{domanski_ligandbook_2017}, (Package ID 2486). 
%%%%%
\subsubsection{Classical force field parameters}
Force field parameters were defined similar to our previous work \cite{yazdanyar_adsorption_2018}. Water was modeled using the SPC/E model \cite{berendsen_missing_1987}. Rutile and its interactions with water were described using the parameterization of Predota \etal. \cite{predota_electric_2004}. The force field parameters for the Tris molecule were taken from Ligandbook \cite{domanski_ligandbook_2017} (Package ID 2486), parameterized using the OPLS-AA force field set \cite{jorgensen_development_1996}. The DL\_FIELD code \cite{yong_descriptions_2016} (v4.3) was used to convert these parameters to those usable by the DL\_POLY MD code \cite{smith_dl_2012}.

Rutile-Tris and water-Tris cross-interaction parameters were parameterized using the Lorentz-Berthelot mixing rules \cite{allen_computer_1989}. A cutoff distance of 12 Å was applied to the short-range van der Waals interactions, as well as the real-space part of the electrostatic interactions. The Ewald summation with a precision of 10\textsuperscript{-6} was used for the long-range electrostatic interactions.
%%%%%
\subsubsection{Simulation details}
The Nosé-Hoover thermostat algorithm, with a relaxation time of 0.5 ps, was used as the thermostat. Under the NVT ensemble, the temperature was controlled to be 37 °C. A time step of 0.7 ps was used for the integration of the equations of motion using the Verlet leapfrog integration algorithm. The simulation box was first equilibrated by running molecular dynamics, in DL\_Classic v1.9 \cite{smith_dl_2012}, for 280 ps. During this step, the central carbon of Tris was kept at a distance of 15 Å from the rutile surface.

In order to estimate the free energy associated with the adsorption of the Tris molecule on the surface, well-tempered metadynamics \cite{barducci_well-tempered_2008} were employed, using PLUMED v2.2.2 \cite{tribello_plumed_2014}. The perpendicular distance of the nitrogen atom of Tris from the surface was considered as the collective variable (CV), referred to hereafter as the distance CV. The CV was limited to a maximal distance of 12 Å from the surface in the z direction. The average position of the oxygen atoms of the surface hydroxyl groups, in the z direction, was defined as the position of the rutile surface. In order to limit the sampling space, the center of mass of Tris was confined within a radius of 5 Å around the surface charge point, on the xy plane.

For well-tempered metadynamics, Gaussian hills with an initial height of 1 kJ.mol\textsuperscript{-1} and a width of 0.5 Å, with a bias factor of 10, were deposited every 1.05 ps. Sampling was performed for 110 ns and trajectories were stored every 10.5 ps. Only one titanium atom in the center of the rutile slab was kept fixed in order to prevent lateral movements of the slab. All other atoms were allowed to move. 
%%%%%
\section{Results and Discussion}
\subsection{Interaction of Tris with rutile}
\begin{figure}[htbp]
\begin{center}
% \fbox{
%   \rule{0pt}{2in} \rule{0.9\linewidth}{0pt}}
\includegraphics[width=\linewidth]{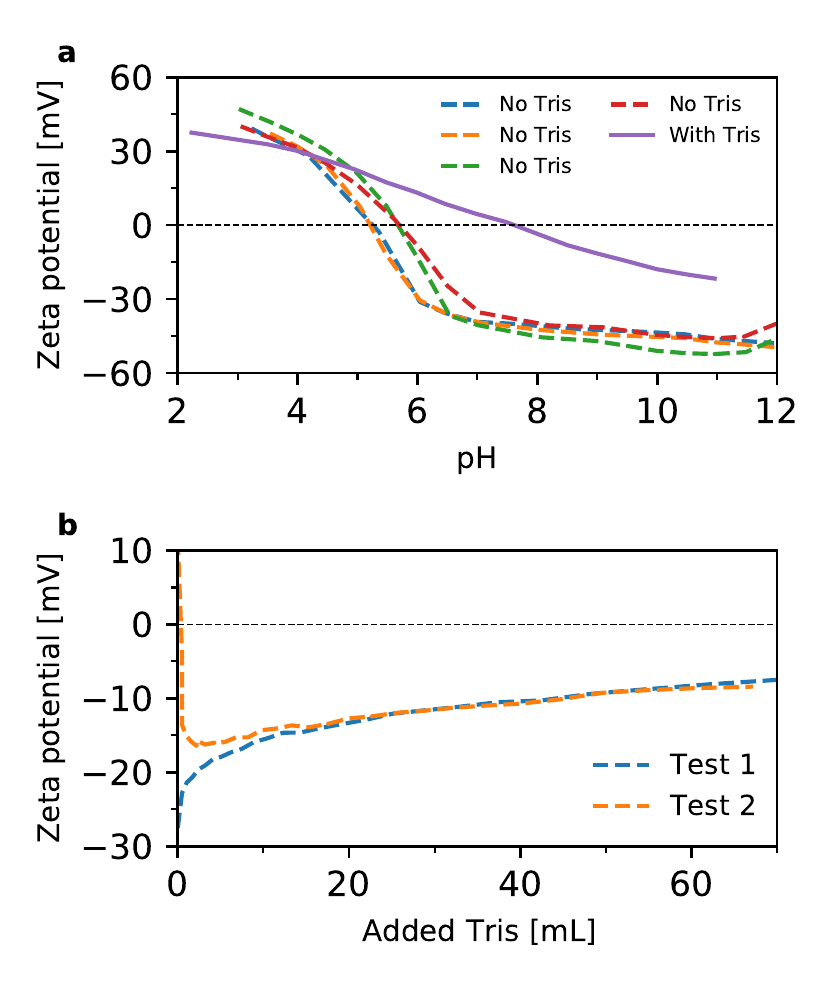}
\end{center}

\caption{a) Zeta potential of rutile as a function of pH in the absence and presence of Tris, showing reproducibility of zeta potential measurements and b) Evolution of the zeta potential upon titration with Tris.}

\label{fig:1}
\end{figure}

\noindent To understand how Tris affects the charge state of the rutile surface, the zeta potential was measured as a function of pH, in presence and absence of Tris (Figure \ref{fig:1}-a). In absence of Tris, the isoelectric point was observed to be 5.46 $\pm$ 0.26, in agreement with previous work \cite{bullard_orientation_2006,parks_isoelectric_1965}. With a negative zeta potential, the rutile surface has a net negative charge for pH values higher than 5.46 $\pm$ 0.26. The rutile suspension, before the addition of Tris, was stable at pH 7.4 with a zeta potential of -38 mV. Nevertheless, a pH of 8.5, with a slightly more negative zeta potential was used for other tests without Tris, to assure stability of the suspension.

Upon titration with Tris, a shift of +2.16 $\pm$ 0.26 is observed in the isoelectric point of rutile. Since the Tris molecule has a net charge of +1 e, the shift towards higher values reveals the adsorption of this molecule on the rutile surface, decreasing the net negative charge density on the surface.
%%%%%%%%%%%%%%%%%%%%%%%
% figure 

%%%%%%%%%%%%%%%%%%%%%%%
Upon titration of the rutile powder with Tris, the zeta potential of the suspension was measured and is presented in Figure \ref{fig:1}-b. The difference in zeta potential between the two curves at the beginning can be explained by the variations of pH. Indeed, the blue curve has a starting pH of 5.69 while the red one starts at a pH of 7.65. After adding approximately 10 mL of Tris, the pH becomes stable at 7.3 and the two curves are superimposed thereafter. The decrease in the magnitude of zeta potential for the two curves after addition of 10 mL of Tris, shows that the positively charged Tris is adsorbing on the rutile surface. This is in agreement with the increase of the isoelectric point when the rutile powder is buffered with Tris (Figure \ref{fig:1}-a) and further agrees with FTIR results of Loreto \etal. indicating a strong adsorption of Tris on the surface of a mesoporous TiO\textsubscript{2} \cite{loreto_effect_2017}.  

\begin{figure}[htbp]
\begin{center}
% \fbox{
%   \rule{0pt}{2in} \rule{0.9\linewidth}{0pt}}
\includegraphics[width=\linewidth]{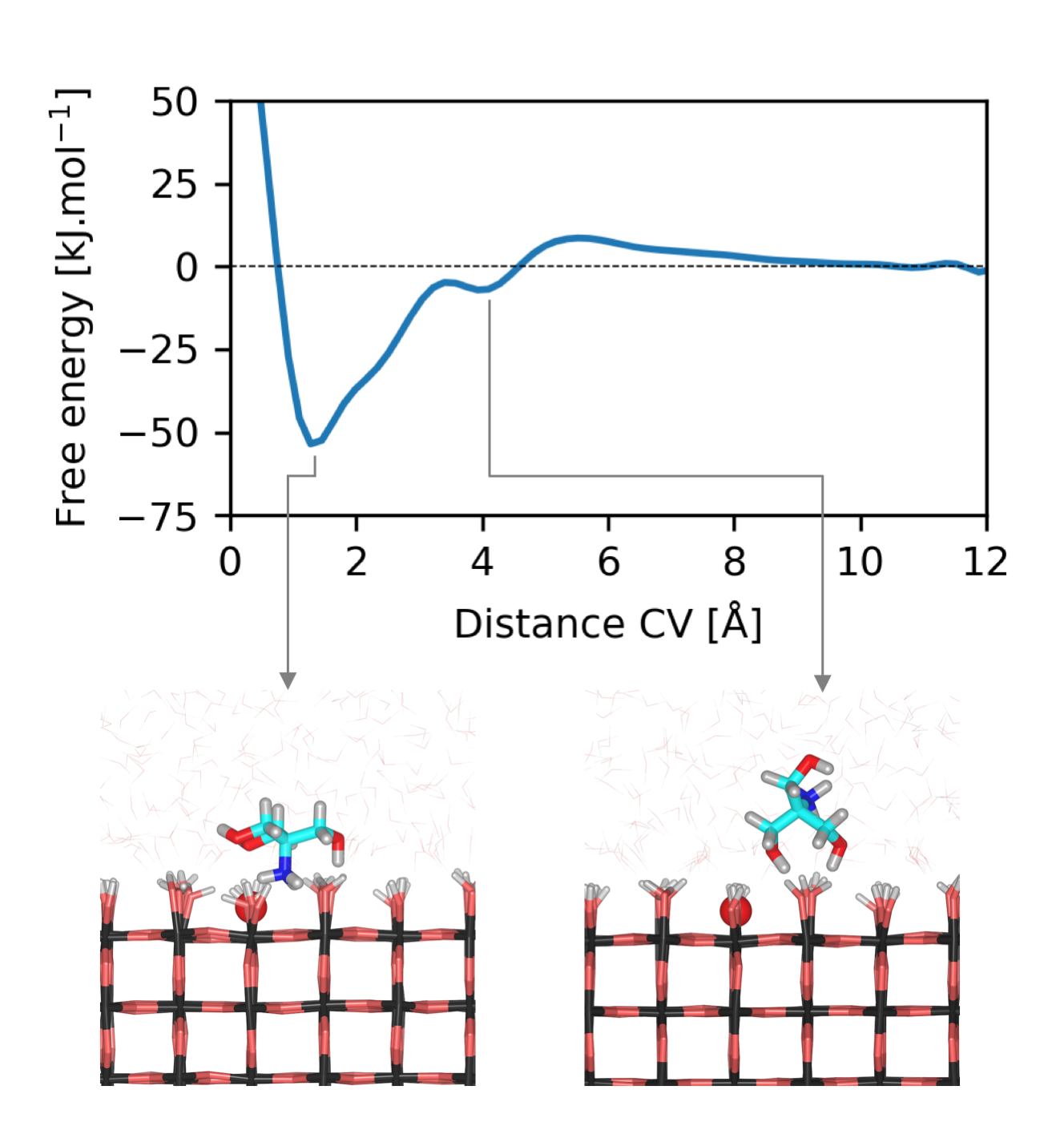}
\end{center}
   \caption{Free energy profile of adsorption of Tris on the rutile surface, as a function of the distance CV. The adsorption conformation of Tris on the surface at the global and local minima are shown (Ti: dark grey, O: red, H: white, C: cyan, N: blue). The surface charge point is shown larger compared to other O atoms of the rutile slab.}
\label{fig:2}
\end{figure}

We also investigated the affinity of the Tris molecule for the rutile surface with the help of computational tools. Here, we used well-tempered metadynamics simulations to obtain the free energy profile of the approach of Tris to the rutile surface. The free energy profiles at the end of the simulation time of 110 ns is shown in Figure \ref{fig:2}, with the distance CV being biased. With the increase in the sampling time, the free energy profiles converge (Figure S2). A global energy minimum and a local energy minimum are observed at distances of 1.27 Å and 3.92 Å from the rutile surface, respectively. We consider the average free energy for distances between 8 to 10 Å from the rutile surface as our reference value. With this definition, the global energy minimum has a depth of -54.98 kJ.mol\textsuperscript{-1} and the local energy minimum has a depth of -8.54 kJ.mol\textsuperscript{-1}. The variation of the distance CV during the sampling time is shown in Figure S3.

The conformation of Tris in these two energy minima is also shown in Figure \ref{fig:2}. While the adsorption of Tris at the global minimum is driven by its amino group, the interaction of Tris with the surface at the local minimum is due to oxygens from the hydroxymethyl groups orientated towards the surface. The classical force field parameterization used in our simulations, attributes the +1 e charge of Tris to its protonated amino group (NH$_3^+$). Since in the global minimum this group is closest to the rutile surface, which has a net negative charge, we can conclude that coulombic interactions are controlling Tris-rutile interactions to a large extent. As previously shown, rutile is hydrophilic and water has a well-defined density profile perpendicular to the surface \cite{predota_electric_2004}. Two high-density water layers were observed close to the surface at distances of 1.85 and 4.35 Å from the surface. The distance CV, which is defined as the distance of the N atom from the surface, has a value of 1.27 and 3.92 Å at the global minimum and the local minimum, respectively. Therefore, in the global minimum the nitrogen atom is closer to the surface than the first high-density water layer and the adsorption is direct with no intermediate water molecules, while in the local minimum, the N atom lies between the first and second high-density water layers and adsorption is indirect.

The adsorption free energy of 54.98 kJ.mol\textsuperscript{-1} of Tris on the surface shows a high affinity of this molecule for the surface, which is in agreement with the significant change in zeta potential upon Tris adsorption (Figure \ref{fig:1}). Based on our previous work \cite{yazdanyar_adsorption_2018}, we are able to compare the affinity of Tris for the rutile surface with that of amino acids with various side groups. These amino acids are present in human blood plasma \cite{stein_free_1954}. The reported values of the adsorption free energy in that work range from ~3 to ~90 kJ.mol\textsuperscript{-1} for various amino acids. We see that the affinity of Tris for the rutile surface lies in this range and is comparable to those of single amino acids.
%%%%%%%%%%%%%%%%%%%%%%%
% figure 
%%%%%%%%%%%%%%%%%%%%%%%
\subsection{Interaction of amino acids with rutile}
\noindent To evaluate the effect of Tris on the adsorption of other species on the rutile surface, we complemented the atomistic simulations \cite{yazdanyar_adsorption_2018} with zeta potential measurements by titrating with four different amino acids covering the different types of side groups, namely Ala (non-polar), Arg (positively charged), Asp (negatively charged), and Ser (polar).

Before reporting the evolution of the zeta potential of the rutile powder when titrating with amino acids, it is important to note that the zeta potential of the rutile surface may change due to two effects. The first one is due to the adsorption of amino acids on the surface ($\zeta$\textsubscript{AA}), and the second one is due to a variation in pH of the solution ($\zeta$\textsubscript{pH}). Assuming that $\zeta$\textsubscript{pH} and $\zeta$\textsubscript{AA} are independent, we can remove the change in zeta potential due to the pH from the overall measured zeta potential to understand the effect due to the adsorption of amino acids alone ($\zeta$\textsubscript{AA} = $\zeta$ - $\zeta$\textsubscript{pH}). 

\begin{figure}[t]
\begin{center}
% \fbox{
%   \rule{0pt}{2in} \rule{0.9\linewidth}{0pt}}
\includegraphics[width=\linewidth]{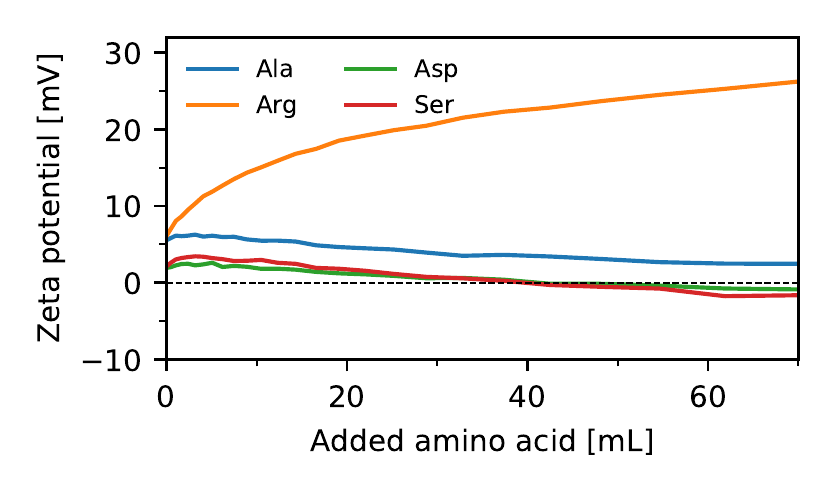}
\end{center}
   \caption{Variation of zeta potential during the titration of the rutile suspension with amino acids at a pH of 8.5. Here only $\zeta$\textsubscript{AA} is shown where $\zeta$\textsubscript{AA} = $\zeta$ - $\zeta$\textsubscript{pH}.}
\label{fig:3}
\end{figure}

To obtain $\zeta$\textsubscript{pH}, linear fitting was performed on the titration of rutile suspensions in the absence of Tris (Figure \ref{fig:1}-a) for pH values between 7 and 11, resulting in $\zeta$\textsubscript{pH} = -2.79 pH - 19.3. The variation of the measured $\zeta$ upon titration of rutile with amino acids is reported in Figure S4; here, we present the results for $\zeta$\textsubscript{AA} (Figure \ref{fig:3}). All the amino acids studied here cause a variation in the zeta potential, indicating that amino acids with all different side groups have adsorbed on the surface, although to different extents. The effect of Arg is the most obvious; Ala and Ser alter the zeta potential to a lesser extent; and Asp has the least effect. Since the rutile surface has a negative charge, the adsorption of the positively charged Arg is expected, and it is indeed observed based on the zeta potential measurements. The adsorption of Ala, Ser and Asp with non-polar, polar and negatively charged side groups, respectively, on the surface is perhaps less expected. However, we observe that they also change the zeta potential and render the surface less positive. This is in good agreement with our previous computational study \cite{yazdanyar_adsorption_2018} where the adsorption of these amino acids on a rutile surface with a net negative charge was shown to be favorable. Moreover, the free energy of adsorption which we previously reported \cite{yazdanyar_adsorption_2018}, also predicted the adsorption energies of Ala and Ser to be very similar to each other (-50.14 and -54.92 kJ.mol\textsuperscript{-1}, respectively), and larger in magnitude than Asp (-32.83 kJ.mol\textsuperscript{-1}). It was shown that for these amino acids, the adsorption on the surface is mainly driven by the backbone of the amino acid.

It should be noted that, as the pH is adjusted to 8.5 but not buffered, small fluctuations in pH occur. These pH fluctuations between the start and the end of the titration are given in Table S2, but are small enough to not greatly influence amino acid adsorption. The influence of these pH fluctuation on the zeta potential has been corrected by the subtraction of $\zeta$\textsubscript{pH}. However, the conformation of the amino acid is pH-dependent and thus the pH fluctuations still affect the adsorption of the amino acids on the rutile surface. Thus, the shape of the zeta potential curves could slightly change if the pH was 8.5 during the whole titration.
%%%%%%%%%%%%%%%%%%%%%%%
% figure 
%%%%%%%%%%%%%%%%%%%%%%%
\subsection{Adsorption of amino acids on rutile in the presence of Tris}
\begin{figure}[b]
\begin{center}
% \fbox{
%   \rule{0pt}{2in} \rule{0.9\linewidth}{0pt}}
\includegraphics[width=\linewidth]{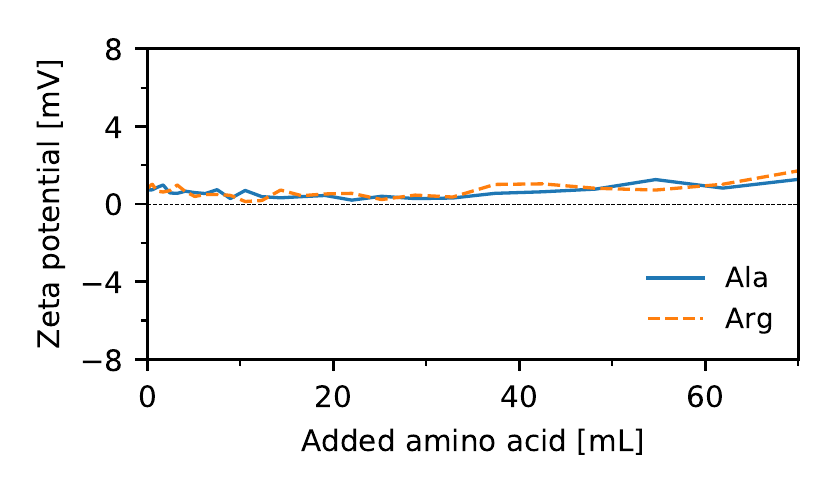}
\end{center}
   \caption{Evolution of zeta potential upon titration of the rutile and Tris suspension with amino acids.}
\label{fig:4}
\end{figure}

\noindent As we observed in the previous sections, titration of the rutile powder with Tris or amino acids always revealed the adsorption of these organic residues on the rutile surface. In a recent study, using surface plasmon resonance, Sipova-Jungova \etal. show that Tris interacts with SAMs (self-assembled monolayers) of negatively charged DNA and alkanethiol groups \cite{sipova-jungova_interaction_2021}. Here, we look at the competitive adsorption of the selected amino acids on rutile with Tris already present, by monitoring the zeta potential, when titrated with either Ala or Arg amino acids. The variation of the zeta potential upon addition of Ala or Arg is shown in Figure \ref{fig:4}. In comparison to Figure \ref{fig:3}, where the addition of amino acids modified the zeta potential by up to ~25 mV depending on the amino acid, in Figure \ref{fig:4}, we observe that the zeta potential is almost constant. This indicates that once Tris is present in the rutile suspension, it prevents adsorption of amino acids on the rutile surface. This is further supported by our computational findings where the affinity of Tris for the rutile surface was computed to be 54.98 kJ.mol\textsuperscript{-1}, which is comparable to that of Ala (50.14 kJ.mol\textsuperscript{-1}) and much larger than the adsorption energy of Arg (3.89 kJ.mol\textsuperscript{-1}) \cite{yazdanyar_adsorption_2018}. As the titration methods are carried out over a short time period (1-2 hrs) there may be modification of the amount of the different molecules at the surface as a function of time, but as the concentration of Tris in the bioactivity test is extremely high at almost 1 M, the likelihood of Tris interfering with surface nucleation of hydroxyapatite is very high. It seems much more judicious to use the natural CO\textsubscript{2} buffer as proposed by Bohner and Lemaitre \cite{bohner_can_2009} a decade ago and recently shown by Zhao \etal. to be more reliable \cite{zhao_comparative_2017}.
%%%%%%%%%%%%%%%%%%%%%%%
% figure 
%%%%%%%%%%%%%%%%%%%%%%%
\subsection{Discussion}
\noindent The current protocol for \textit{in vitro} bioactivity testing of titanium-based biomaterials suggests an ionic solution (Simulated Body Fluid or SBF) for testing, where Tris (tris(hydroxymethyl)aminomethane) is used as the buffering agent \cite{noauthor_iso_2007}. The \textit{in vivo} condition, in which a sample is tested, contains many organic residues. At first glance, one may conclude that in order to have a more representative \textit{in vitro} solution, organic molecules, which can be as simple and small as amino acids, or as complicated as Human Serum Albumin (HSA), should be introduced into the SBF solution. However, our findings for two test amino acids (Ala and Arg) revealed that the adsorption of amino acids is inhibited to a great extent when Tris is present in the solution. One can thus expect unaltered interaction behavior if a Tris buffered SBF solution is prepared and amino acids are added to the solution to represent the organic residues that are present \textit{in vivo}. Our results show that the interaction of amino acids with the rutile surface will be greatly inhibited by Tris, given that the concentration of Tris in the ISO standard \cite{noauthor_iso_2007} is 974 mM, which is about 200 times higher than the concentration of amino acids in blood plasma and about two times higher than the total concentration of proteins in blood plasma \cite{marieb_human_2013}.

However, our computational findings also reveal that the use of Tris as the buffering agent in ISO \cite{noauthor_iso_2007} may not be totally negative. Here, we show that the affinity of Tris for the rutile surface is comparable to those of simple amino acids. Therefore, the presence of Tris in the SBF solution, although as the buffering agent, may play a role very similar to those of amino acids in an \textit{in vivo} setting. The competition between ionic and organic species for the rutile surface, may be mimicked, to some extent, by the presence of Tris in the SBF solution. However, the very high concentration of Tris in SBF may overly hinder the interactions of ionic species with the surface and influence the formation of hydroxyapatite. As an example, the ability of rutile to form apatite was measured in various concentrations of bovine serum albumin by Zhao \etal. \cite{zhao_comparative_2017}, where they reported that the apatite forming ability of the surface is completely inhibited for concentrations higher than 0.075 mM.

In the human body, the partial pressure of CO\textsubscript{2} controls the pH in blood. Recent studies \cite{zhao_comparative_2017,zhao_rapid_2019} have succeeded in using a 5\% CO\textsubscript{2} for \textit{in vitro} testing with SBF instead of using Tris. This allows for a more realistic \textit{in vitro} representation of the \textit{in vivo} conditions. In such a system, one can be certain that interactions are not hindered by the buffering agent in the system and the reliability of results is improved. As already stated, in a Tris-buffered system, amino acids or proteins added to the SBF may not be able to adsorb on the surface or will be in direct competition with Tris. The very high concentration of Tris could also kinetically inhibit the amino acid adsorption on the surface. Thus, the correlation between \textit{in vitro} and \textit{in vivo} tests is weak and indicates that the use of the carbonate buffer approach instead of Tris is the best way forward for more reliable \textit{in vitro} studies. 

\section{Conclusion}
\noindent We investigated the affinity of Tris (tris(hydroxymethyl)aminomethane) used as the buffering agent in the current standard \textit{in vitro} testing protocol for titanium-based biomaterials. In order to do so, we investigated the interaction of Tris with a model rutile surface. Both our experimental and computational findings reveal that Tris interacts strongly with the rutile surface. Tris shifts the surface charge density of rutile (which is negative under physiological conditions), to more positive values. We also investigated the interaction of simple amino acids with the rutile surface via zeta potential measurements, as possible organic residue candidates to be added to the current SBF solution to better represent the \textit{in vivo} test settings. The amino acids studied here (Ala, Arg, Asp and Ser), covering non-polar, polar and charged side groups, all modified the rutile zeta potential. Our results show that the interactions of amino acids (Ala and Arg) were greatly weakened by the presence of Tris. This can be due to two reasons. The first reason is that the free energy of adsorption of Tris on the surface was found to be in the same range as of amino acids. Therefore, amino acids may not be able to easily displace Tris on the rutile surface. The second reason is the high concentration of Tris in the SBF solutions, which could kinetically limit the adsorption of the amino acids. Therefore, we conclude that under the current ISO protocol, amino acids are not suitable to represent the organic residues of \textit{in vivo} settings.

Our results thus reveal a strong interaction of Tris with our model rutile surface. Results of Tris-buffered \textit{in vitro} tests, therefore, should be interpreted with care, and the possibility of a strong interaction of Tris with surfaces may be one of the main causes of the high number of false negative and false positive results which are observed in bioactivity testing of titanium-based biomaterials. The use of the natural buffer, CO\textsubscript{2}, is a far more reliable way of buffering \textit{in vitro} tests and will allow clear interpretation of other molecules and ions with surfaces of biomedical interest.

\section*{Acknowledgement}
\noindent This work was realized by the support of the Swiss National Science Foundation [SNSF BioCompSurf 513468]. AYY acknowledges the CPU time on EPFL clusters provided through the Scientific IT and Application Support Center (SCITAS). The authors acknowledge Weitian Zhao for his experimental help and Carlos Morais for the experimental training.

\section*{Competing Interests}
\noindent None.

%------------------------------------------------------------------------
{\small
\bibliographystyle{ieeetr}
\bibliography{egbib}
}

\end{document}

%%%%%%%%% TITLE
\title{Supporting Information  for\\
\bigskip
Interactions of Tris with rutile surfaces and consequences for \textit{in vitro} bioactivity testing}

\author{\underline{Azade YazdanYar}\textsuperscript{1 %\href{ayazdan@icp.uni-stuttgart.de}{\textbardbl}
\thanks{Present address: Institute for Computational Physics, Universität Stuttgart, Allmandring 3, 70569 Stuttgart, Germany; ayazdan@icp.uni-stuttgart.de} }, 
Léa Buswell\textsuperscript{1}, 
Delphin Pantaloni\textsuperscript{1 \thanks{Present address: Université de Bretagne-Sud, IRDL, CNRS UMR 6027, BP 92116, 56321 Lorient Cedex, France}}, 
Ulrich Aschauer\textsuperscript{2}, 
Paul Bowen\vspace{4mm}\textsuperscript{1}\\

\small \textsuperscript{1}Department of Materials Science and Engineering, École Polytechnique Fédérale de Lausanne (EPFL), \\
\small Route Cantonale, Lausanne 1015, Switzerland\\
\small \textsuperscript{2}Department of Chemistry and Biochemistry, University of Bern, Freiestrasse 3, Bern 3012, Switzerland
}

\maketitle

\renewcommand{\thefigure}{S\arabic{figure}}
\renewcommand{\thetable}{S\arabic{table}}
\renewcommand\thepage{S\arabic{page}}
\floatsetup[table]{capposition=top}
%%%%%%%%%%%%%%%%%%%%%%%
% figure 

\begin{figure*}[h]
\begin{center}
% \fbox{
%   \rule{0pt}{2in} \rule{0.9\linewidth}{0pt}}
\includegraphics[width=0.7\textwidth]{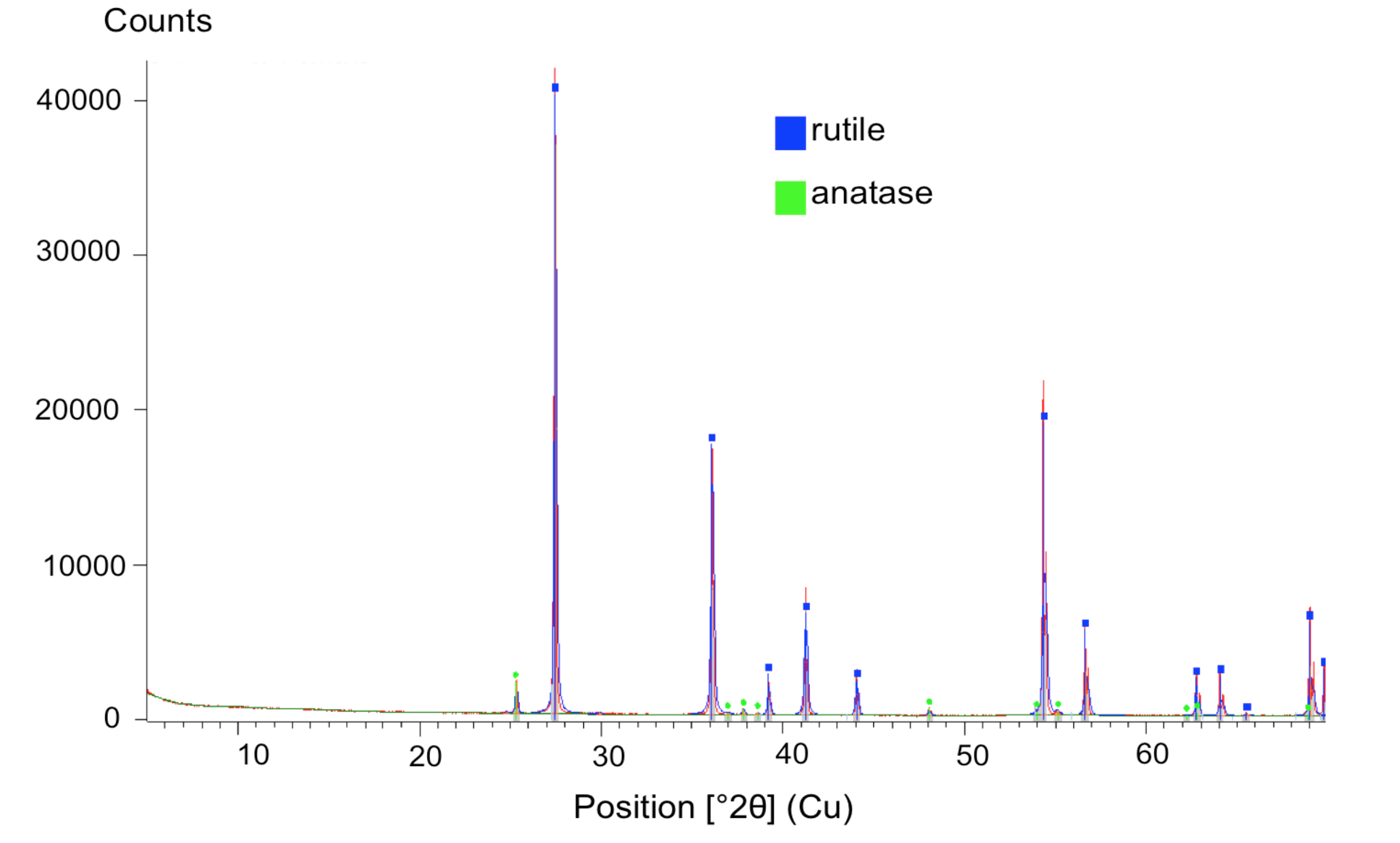}
% }
\end{center}
\begin{center}
\caption{The XRD pattern of the rutile powder.}
\end{center}
\label{fig:s1}
\end{figure*}
%%%%%%%%%%%%%%%%%%%%%%%
\begin{table}[h]
\caption{Details of the purchased products.}
\begin{center}
\begin{tabular}{ l | c | c | c }
\small \\[-1em]
\textbf{Product} & 
\textbf{Supplier} & 
\textbf{Assay} &
\textbf{Reference}\\
\hline \\[-1em]
Rutile powder & Sigma-Aldrich & - & 204757\\
NaCl & ACROS organics & - & 447302500\\
Tris & EDM Millipore & - & 1.08382.0500\\
HCl & Carlo Erba Reagents & - & 404097000\\
NaOH & ACROS organics & - & 124260010\\
Alanine & ACROS organics & 99\% & 102830250\\
Serine & ACROS organics & 99\% & 132660250\\
Arginine & Sigma Aldrich & 99.5\% & 11009-25-G-F\\
Aspartic acid & Sigma Aldrich & 99\% & A8949-25G\\
Alanine & ACROS organics & 99\% & 102830250\\

\end{tabular}
\end{center}

\label{tab:table_s1}
\end{table}
%%%%%%%%%%%%%%%%%%%%%%%
% figure 

\begin{figure}[h]
\begin{center}
% \fbox{
%   \rule{0pt}{2in} \rule{0.9\linewidth}{0pt}}
\includegraphics[width=0.6\linewidth]{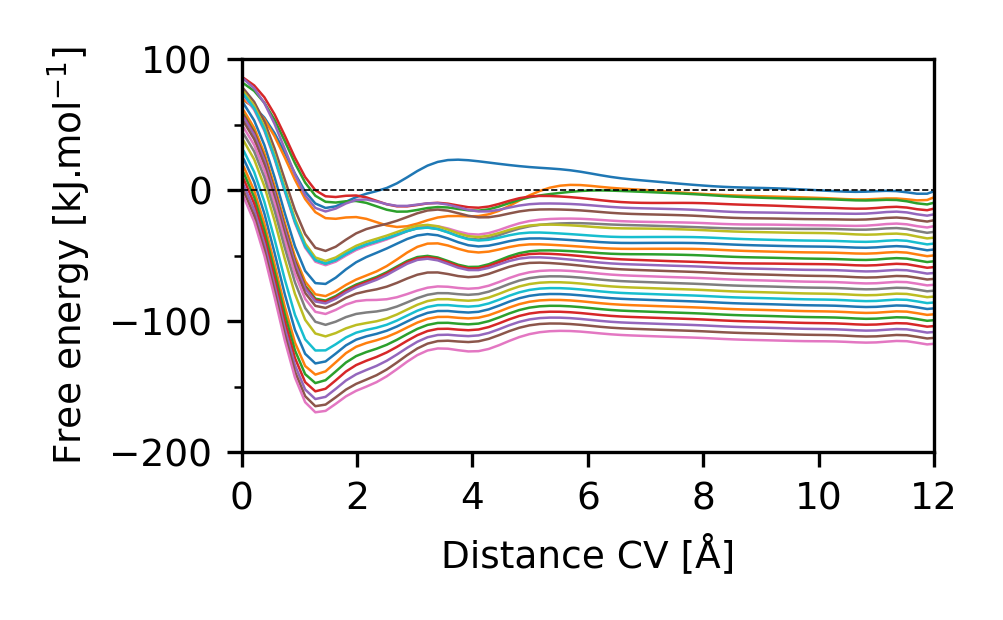}
\end{center}
\caption{Free energy profiles of adsorption of Tris on the rutile surface, as a function of the distance CV. Free energy profiles are plotted over intervals of 4 ns, with a downward offset in order to show the convergence of the enhanced sampling over time.}

\label{fig:s2}
\end{figure}
%%%%%%%%%%%%%%%%%%%%%%%
%%%%%%%%%%%%%%%%%%%%%%%
% figure 
\begin{figure}[h]
\begin{center}
% \fbox{
%   \rule{0pt}{2in} \rule{0.9\linewidth}{0pt}}
\includegraphics[width=0.6\linewidth]{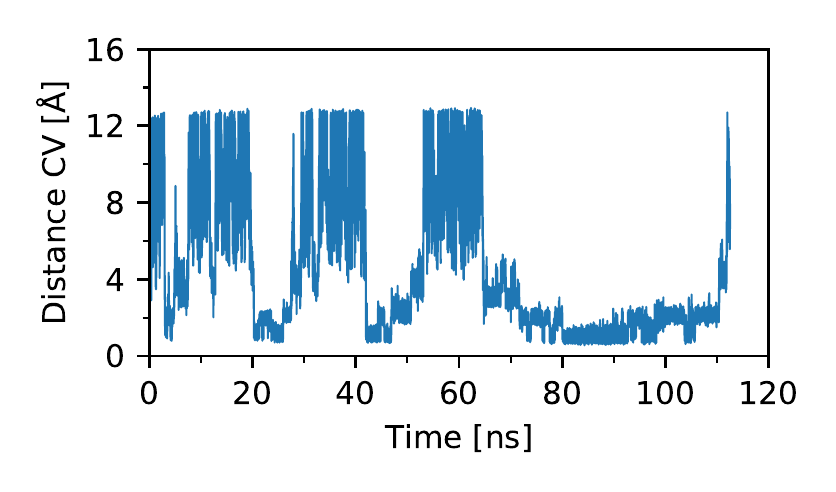}
\end{center}
\caption{Variation of the biased collective variable (distance of the nitrogen atom from the surface)  over the simulation time.}
\label{fig:s3}
\end{figure}
%%%%%%%%%%%%%%%%%%%%%%%
%%%%%%%%%%%%%%%%%%%%%%%
% figure 

\begin{figure}[h]
\begin{center}
% \fbox{
%   \rule{0pt}{2in} \rule{0.9\linewidth}{0pt}}
\includegraphics[width=0.6\linewidth]{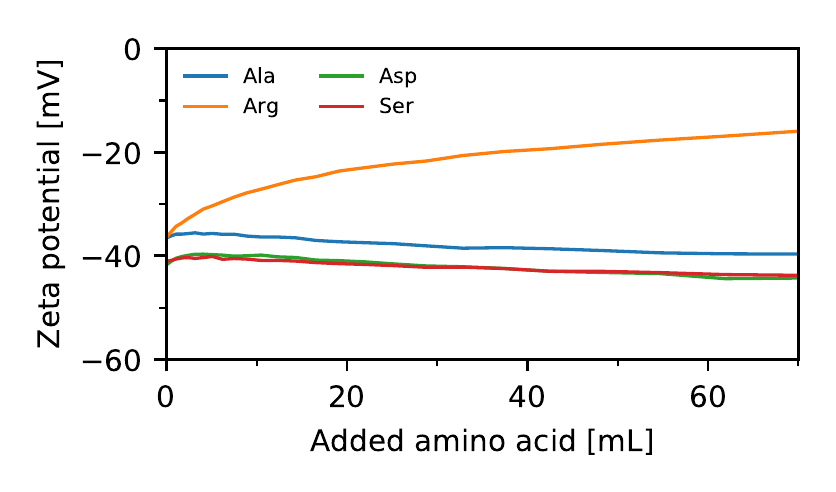}
\end{center}

\caption{The total zeta potential of the rutile powder during titration with amino acids.}

\label{fig:s4}
\end{figure}
%%%%%%%%%%%%%%%%%%%%%%%

\begin{table}[h]
\caption{
pH of the suspension at the start and the end of the titration for each amino acid.}
\begin{center}

\begin{tabular}{ l | c | c }
\small \\[-1em]
\textbf{Amino acid} & \textbf{pH\textsubscript{start}} & \textbf{pH\textsubscript{end}}\\
\hline \\[-1em]
Alanine & 8.15 & 8.17\\
Arginine & 8.31 & 8.18\\
Aspartic acid & 8.47 & 8.44\\
Serine & 8.79 & 8.36\\
\end{tabular}
\end{center}

\label{tab:table_s2}
\end{table}